\documentclass{article}
\usepackage{amsmath,amsfonts,amssymb,array,epsfig,hyperref}
\title{Limit on the mass of a long-lived or stable gluino}
\author{G.~R.~Farrar$^a$, R.~Mackeprang$^b$, D.~Milstead$^c$, J.~P.~Roberts$^a$}
\date{\today}
\newcommand{\spi}{\ensuremath{\tilde{\pi}}}
\newcommand{\spin}{\ensuremath{\tilde{\pi}^0}}
\newcommand{\spim}{\ensuremath{\tilde{\pi}^-}}
\newcommand{\spip}{\ensuremath{\tilde{\pi}^+}}
\newcommand{\slam}{\ensuremath{S^0}}
\newcommand{\ska}{\ensuremath{\tilde{K}}}
\newcommand{\skan}{\ensuremath{\tilde{K}^0}}
\newcommand{\skam}{\ensuremath{\tilde{K}^-}}
\newcommand{\skap}{\ensuremath{\tilde{K}^+}}
\newcommand{\skanb}{\ensuremath{\bar{\tilde{K}}^0}}
\newcommand{\pin}{\ensuremath{\pi^0}}
\newcommand{\pim}{\ensuremath{\pi^-}}
\newcommand{\pip}{\ensuremath{\pi^+}}
\newcommand{\kn}{\ensuremath{K^0}}
\newcommand{\knb}{\ensuremath{\bar{K}^0}}
\newcommand{\km}{\ensuremath{K^-}} \newcommand{\kp}{\ensuremath{K^+}}
\newcommand{\spro}{\ensuremath{\tilde{p}}}
\newcommand{\sneu}{\ensuremath{\tilde{n}}}
\newcommand{\delp}{\ensuremath{\tilde{\Delta}^+}}
\newcommand{\deln}{\ensuremath{\tilde{\Delta}^0}}

\begin{document}
\maketitle
%\\[5mm]
{\small\it
$^a$Center for Cosmology and Particle Physics and Department of Physics, New York University, % New York, NY 10003,
USA}
\\%[5mm]
{\small\it
$^c$Discovery Center for Particle Physics, Niels Bohr Institute, Blegdamsvej 17, Copenhagen, Denmark}
\\%[5mm]
{\small\it
$^b$Fysikum, Stockholms Universitet, Stockholm, Sweden}\\[1mm]
\begin{abstract}
  We reinterpret the generic CDF charged massive particle limit to obtain a limit on the mass of a stable or long-lived gluino.  Various sources of uncertainty are examined.  The $R$-hadron spectrum and scattering cross sections are modeled based on known low-energy hadron physics and the resultant uncertainties are quantified and found to be small compared to uncertainties from the scale dependence of the NLO pQCD production cross sections.  The largest uncertainty in the limit comes from the unknown squark mass:  when the squark -- gluino mass splitting is small, we obtain a gluino mass limit of 407 GeV, while in the limit of heavy squarks the gluino mass limit is 397 GeV.   For arbitrary (degenerate) squark masses, we obtain a lower limit of 322 GeV on the gluino mass.  These limits apply for any gluino lifetime longer than $\sim 30$ ns, and are the most stringent limits for such a long-lived or stable gluino.
\end{abstract}
\setcounter{footnote}{0}
\section{Introduction}

The observation of exotic stable massive particles (SMPs) which can be detected by their
interactions in a detector would be of fundamental significance. SMPs
are features of a number of scenarios of physics beyond the Standard
Model, such as theories of supersymmetry (SUSY) and extra
dimensions~\cite{Fairbairn:2006gg}. Searches have therefore been
carried out at colliders, in cosmic rays and
matter~\cite{Fairbairn:2006gg,Amsler:2008zzb}. Collider searches for
coloured SMPs -- $R$-hadrons in the context of SUSY -- present
an additional challenge in interpreting the experimental observations,
owing to uncertainties in the mass spectra and scattering of the colour-singlet
$R$-hadrons. This is of particular concern for
$R$-hadrons consisting of a heavy colour octet combined with a gluon
or $q \bar{q}$ in a colour-octet state, since these systems have no direct analog among Standard
Model hadrons~\cite{Farrar:1984gk}.  Consequently, there have been no
dedicated searches for stable gluino $R$-hadrons since the LEP
era~\cite{Heister:2003hc,Abdallah:2002qi}.  In this paper, we outline an approach to modeling the mass spectrum and the scattering of gluino-based $R$-hadrons.  This provides the theoretical and phenomenological foundation for a new method to search for long-lived or stable massive gluinos.  Using our method, we reinterpret a recent CDF limit~\cite{Aaltonen:2009kea} to
give a new lower limit on the mass of a stable or long-lived gluino.  

There have been dedicated searches for {\it unstable} massive gluinos since the LEP era. D\O ~performed a search for unstable gluinos that stop in the detector and decay \cite{Abazov:2007ht}. While the present paper was in review, CMS presented results of an analogous search at the LHC \cite{Collaboration:2010uf}.  Searches for decaying gluinos are highly dependent on the lifetime of the gluino, whereas our approach only requires that the gluino have a lifetime long enough that it traverses the detector without decaying in flight, $\tau>30$~ns, a condition satisfied by all the models considered in stopped gluino searches.  Thus our limits can be directly compared to those studies but have a larger range of applicability.  For brevity below, we call a particle stable if it does not decay during its traversal of a detector.

There are two main approaches to search for non-decaying strongly interacting
SMPs at a collider\footnote{We only consider strongly interacting SMPs in the remainder of this work, and denote them hereafter as simply SMPs for brevity.}. One strategy exploits the expected anomalous
energy loss of a SMP as it propagates through an inner tracking system
next to the beam-line, while the other relies on the SMP-speed ($\beta
\ll 1$) and possible penetrating behaviour whereby a SMP can be
measured as a slow-moving object in an outer muon tracking
system. Typically, experiments at
LEP~\cite{Heister:2003hc,Abdallah:2002qi} and HERA~\cite{Aktas:2004pq}
exploited the former, while the Tevatron
experiments~\cite{Aaltonen:2009kea,Abazov:2008qu} relied on the latter
signature. However, since $R$-hadrons interact hadronically in dense calorimeter
material positioned between the aforementioned tracking
systems~\cite{Kraan:2004tz,deBoer:2007ii,Mackeprang:2009ad}, the
$R$-hadron charge can change between the inner detector and when it
reaches the outer muon system. The charge of the $R$-hadron in the
outer muon system depends both on the scattering mechanism and the
$R$-hadron mass hierarchies. A comprehensive search for $R$-hadrons at
a collider must consider $R$-hadron models in which the $R$-hadron
{\em i)} is dominantly neutral throughout its passage in a detector,
{\em ii)} can be charged in the inner system but is dominantly neutral
in the outer tracking systems, and {\em iii)} can be charged in both
the inner and outer trackers. Since the current strongly-interacting
SMP-limits from the Tevatron assume the latter scenario, we provide
here a gluino mass limit applicable to models which predict the
muon-like signature {\em iii)}.

This paper is organised as follows. First we outline expectations for
the gluino-containing $R$-hadron mass spectra.  This is followed by a
description of the simulation of $R$-hadron scattering in
matter. Using the material composition of the CDF detector and fixed
order QCD-models, the proportions of gluino $R$-hadrons which would
have passed the CDF SMP-selections are then estimated and mass limits
extracted.

\section{Spectrum of R-hadron masses}\label{sec:spec}

To consider the experimental signatures of $R$-hadrons in a detector
we need to postulate a mass spectrum for the $R$-mesons and the
$R$-baryons.  The masses of the $R$-hadrons have been calculated using
the MIT bag model \cite{DeGrand:1975cf} for $R$-mesons
\cite{Chanowitz} and $R$-baryons \cite{Buccella}\footnote{Lattice calculations have been performed
for $R$-mesons and gluino-balls, e.g., \cite{UKQCD-Rmasses}, but the bag-model results are consistent with the lattice calculations and easier to use.}. The mass splittings within a given $R$-hadron
multiplet are governed by the same QCD interactions that are
responsible for the mass splittings of ordinary mesons and baryons.
These are surprisingly well-described by an effective
one-gluon-exchange interaction, and many unknown parameters for the
$R$-hadron mass splittings can be fixed by splittings in the analogous
ordinary hadrons.  To the extent that parameters are determined
phenomenologically, splittings should be fairly model-independent.

An important feature of the $R$-baryon mass spectrum, first pointed out
by Farrar \cite{Farrar:1984gk}, is that the lightest $R$-baryon is the
neutral flavor-singlet $u \, d \, s \,\tilde{g}$, denoted $S^0$.  Although the
constituent quarks are the same, this is a distinct state from the
flavor-octet $\tilde{\Lambda}^0$.  Due to the particularly strong
hyperfine attraction in the flavor-singlet channel, the $S^0$ is
lighter than the $R$-proton. There is no flavor-singlet ordinary
baryon analog of the $S^0$, due to Fermi statistics when the quarks
form a colour-singlet.   As noted above, the QCD hyperfine interactions that govern the R-baryon spectrum are well-constrained by the spectrum of ordinary hadrons and our knowledge of low-energy QCD, making it highly unlikely that the $S^0$ would not be the lightest state.  The signal in our analysis would be bigger for a given gluino mass if the lightest R-baryon were charged, so doing the analysis with the most plausible mass spectrum (the $S^0$ being lightest) gives the most conservative gluino mass limits.

The $R$-meson spectrum is important for our analysis, because the
initial $R$-hadrons resulting from hadronization of a gluino are
predominantly $R$-mesons.  $R$-mesons are formed by the combination of
the gluino with a colour octet $q\overline{q}$ pair. We must also
consider the gluino-ball $\tilde{g}g$ on a similar footing as it mixes
directly with the flavor singlet $R$-meson states.  The states can be
broken down by first considering the $q\overline{q}$ state and then
combining this with the spin-$\frac{1}{2}$ $\tilde{g}$. The combination of the
$J^{C}=0^{-}$ $q\overline{q}$ state with the $1/2^-$ gluino
gives a $1/2^+$ $R$-meson. The combination of the $1^+$ $q\overline{q}$
state with the gluino gives both a $1/2^-$ nonet and a $3/2^+$ nonet.
It is sufficient to focus only on the lightest states so we simplify
the meson sector to include the two lightest nonets with $J=1/2$.  In
\cite{Chanowitz} it is shown that as $m_{\tilde{g}}\rightarrow \infty
$, the mass splitting between states becomes constant: $m_{\tilde
  K}-m_{\tilde \rho}=130$~MeV, $m_{\tilde \pi}-m_{\tilde
  \rho}=40$~MeV. It is important that the $R$-pions ($1/2^+$) are {\em
  not} significantly lower in mass than other non-strange $R$-mesons,
unlike the ordinary pions, because they are not pseudo-Goldstone
bosons.  Thus we take the $R$-meson masses to be split only by the
strange quark content\footnote{ The $R$-pions are actually expected to
  be 40~MeV heavier than the $R$-rho according to \cite{Chanowitz}, but this is
  an unnecessary level of refinement and we ignore it here.}. With
two light nonets and a gluino-ball we have 19 states: 10 degenerate
non-strange $R$-mesons, a gluino-ball we take to be degenerate, and a
heavier set of 8 $R$-kaons. We take the mass splitting between these
states to be $130$~MeV.  For ease of reference we refer to all
non-strange $R$-mesons as $\tilde \pi$ from here on, and all strange
$R$-mesons as $\tilde K$.

\begin{table}[h!]
\begin{center}
\begin{tabular}{|lc|}
\hline
$m_{\tilde{K}}-m_{\tilde{\pi}}$ & $130$~MeV \\
\hline
\end{tabular}
\end{center}
\caption{The mass splitting of the $R$-meson states stable against strong decays from \cite{Chanowitz}.\label{tab:RMesonMasses}}
\end{table}

In the $R$-baryon sector the gluino combines with a $qqq$ colour
octet. The $qqq$ state can have either $J=1/2$ or $J=3/2$; 
combined with the spin-$1/2$ gluino this gives a set of states with
$J=0,1,2$. Fermi statistics requires that the $J=3/2$ $qqq$ state
be a flavour octet, while the $J=1/2$ state can be a flavour singlet,
octet or decuplet. Buccella et al \cite{Buccella} showed that for
gluino masses above $1$~GeV, the only $R$-baryon states that are stable
against strong decays are the $J=0$ flavour singlet $\tilde{g}uds$ ($S^0$), the
flavour octet with $J=0$ and the lighter of the two flavour octets
with $J=1$.   

We can estimate the weak-decay lifetime for $R_N \rightarrow S^0 + \pi^\pm$ to be
\begin{equation}
\tau_{R-N} \approx 1.6 \times 10^{-6} \,{\rm sec} \left( \frac{M_{\tilde{g}}}{100 \, \rm GeV} \right)^2,
\end{equation}
by scaling the rate for $\Lambda \rightarrow N + \pi$ with particle masses as appropriate for 2-body decay kinematics.
Thus, we can ignore weak decays in the rest of our analysis.

\begin{table}[h!]
\begin{center}
\begin{tabular}{|lc|}
\hline
$m_{8_1}-m_{1_0}$ & $250$~MeV\\
$m_{8_0}-m_{8_1}$ & $130$~MeV\\
\hline
\end{tabular}
\end{center}
\caption{The mass splitting of the $R$-baryon states stable against strong decays from \cite{Buccella}.\label{tab:RBaryonMasses}}
\end{table}

The splittings between the states are summarised in Table
\ref{tab:RBaryonMasses}\footnote{There is some uncertainty in the mass
  splitting that arises from the value of the zero point energy taken
  in the bag model calculation. We take the values in Table
  \ref{tab:RBaryonMasses} for our model and then vary the splitting to
  check that our results are robust with respect to the
  splittings.}. In practice we can ignore many of the $R$-baryons. Any
singly strange state in the octet can decay strongly to the flavour
singlet and so should not be considered to be quasi-stable. Moreover
because the $R$-baryons are created through interactions of the form
$\tilde{\pi}+N\rightarrow \tilde{N}+\pi$ we can ignore the doubly
strange states: there is no way of creating a doubly strange state
from the interactions of an $R$-meson with the nucleons in the
detector. We designate the singlet state as $S^0$, the spin-0 states
as $\tilde{n},~\tilde{p}$ and the spin-1 states as
$\tilde{\Delta}^0,~\tilde{\Delta}^+$.

The mass calculations of \cite{Chanowitz} and \cite{Buccella} were
performed using several choices of bag model parameters.  The primary effect of
the different parameter choices is to alter the overall scale of the
masses. Therefore we keep the interstate splitting within the meson
and baryon sectors. We take the mass splitting between the meson and
baryon sectors to be equivalent to the mass of the extra valence quark
which we set to $330$~MeV. In our final results we allow for this to
be varied and find that this splitting has little effect on the
overall conclusions.

\begin{table}[!ht]
\begin{center}
\begin{tabular}{|lc|}
\hline
\multicolumn{2}{|l|}{$R$-baryons} \\
\hline
$R_{uds}$: $S^0$ & $m_{\tilde{g}}+280$~MeV \\
$R_{8-0}$: $\tilde{\Delta}^{+/0}$ & $m_{\tilde{g}}+530$~MeV \\
$R_{8-0}$: $\tilde{n},~\tilde{p}$ & $m_{\tilde{g}}+660$~MeV \\
\hline
\multicolumn{2}{|l|}{$R$-mesons} \\
\hline
$\tilde{\pi}$ & $m_{\tilde{g}}+330$~MeV \\
$\tilde{K}$ & $m_{\tilde{g}}+460$~MeV \\
\hline
\end{tabular}
\end{center}
\caption{
 Masses used for the $R$-hadrons.\label{tab:RHadronMasses}
}
\end{table}

Finally, we must set the overall mass normalisation with respect to the
gluino mass. This value is sensitive to the details of the bag model
calculations. We follow Buccella et al \cite{Buccella}, setting
$m(\tilde{n})=m(\tilde{g})+660$~MeV. This fixes the overall mass scale
and gives the spectrum listed in Table~\ref{tab:RHadronMasses}. In our
final results we allow this overall scale to vary and include the
effect in our estimation of the theoretical uncertainty

\section{Propagation of R-hadrons through a detector}\label{sec:prop}
Following production in a hard collision, a gluino would fragment
dominantly into either an $R$-meson or gluino-gluon state.  This
particle would then propagate through the detector. In order of increasing
distance from the interaction point, a typical multipurpose experiment
at a collider comprises an inner tracking system, a calorimeter and
an outer set of muon chambers. If charged, the momentum of the $R$-hadron would
be measured in the inner and outer tracking systems. Hadronic and
electromagnetic energy loss of the $R$-hadron would be recorded in the
calorimeter.

Hadronic interactions of $R$-hadrons occur predominantly due to the interactions
between target nucleons and the light quark system accompanying the gluino, with the much
more massive gluino acting as a spectator~\cite{Kraan:2005ji}. The
full list of hadronic processes we consider is given in
Appendix~\ref{app:lists}. For $R$-hadrons produced at the Tevatron the
light quark system typically has $\mathcal{O}$(GeV) kinetic energy,
implying that the interactions with nucleons should resemble low
energy hadronic interactions. In such processes quarks can be
exchanged with the quarks in the target nucleon, which can change the
charge of the $R$-hadron. Another process in
$R$-hadron scattering with nucleons is the conversion of $R$-meson and
gluino-gluon states to baryon states with pion production. Owing to
lightness of the pions, such processes are exothermic and thus
energetically favoured. Furthermore, the absence of pions in the
scattering material renders the reverse process unlikely. In the
scattering model used in this work, an $R$-hadron undergoes on average
$\sim$ 5 hadronic interactions in 2 m of iron~\cite{Mackeprang:2009ad},
which is representative of the material within a typical calorimeter
at a collider experiment. Thus most $R$-hadrons would enter the muon
detector as baryons, predominantly the spin-1 octet states $R_{uud}$
and $R_{udd}$ and the spin-zero singlet state $S^0$ ($R_{uds}$), as
discussed in Sec. \ref{sec:spec}.

\section{Simulation of R-hadrons}\label{sec:sim}

To simulate the $R$-hadron signal in the detector we employ several
computer codes.

\begin{figure}[!ht]
  \centering
  \epsfig{file=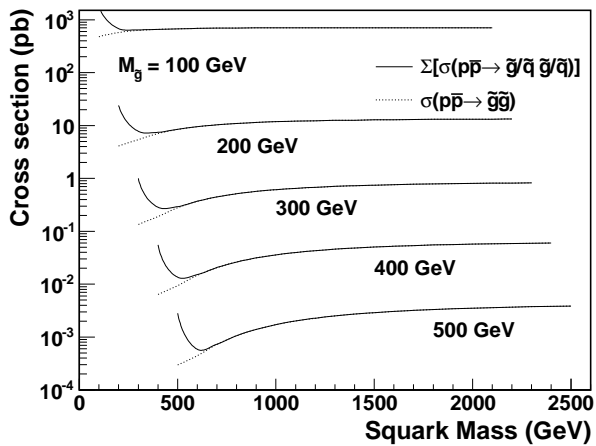,width=8cm}
  \caption{NLO gluino production cross sections at the tevatron for five different gluino masses. The dotted line shows the cross-section for direct production of a pair of gluinos. The solid line shows the total cross-section for the production of any pair of coloured SUSY particles, relevant because squarks decay virtually instantaneously to a gluino. As the squark masses increase the squarks decouple and the two cross-sections converge.}
  \label{fig:massgrid}
\end{figure}

{\bf Gluino production} \\The cross sections for gluino pair
production are calculated using Prospino2.1 \cite{Beenakker:1996ch}
assuming all squarks have a common mass. The cross section
dependence on the squark and gluino masses is shown in Figure
\ref{fig:massgrid} for the Tevatron. Two cross-sections are plotted for each gluino mass. The dotted line shows the direct pair production cross section for two gluinos. The solid line shows the total cross-section for the production of any pair of coloured SUSY states (squark-antisquark, gluino-(anti)squark, and gluino-gluino). As the squark mass increases, squark production falls off and the two cross-sections converge. Any squark produced decays to a quark and gluino, effectively instantaneously\footnote{We ignore such small mass splittings, $\Delta m \ll 1$ GeV, that the squark decay is not effectively instantaneous;  even though with small mass splitting the stop, sbottom and scharm states would be stable, these provide a negligible contribution to the final states.}. We use this to defines three regimes in which we set mass limits:
\begin{enumerate}
\item {\it Decoupled squarks}. For large squark masses the cross-section for the production of two gluinos becomes independent of the squark mass and is just the cross-section for the direct production of a gluino pair. This can be compared directly with the split SUSY scenarios commonly used in gluino searches.
\item {\it Light squarks}. In the other extreme, that squarks are light (but heavier than the gluino as required for a long-lived gluino), an important source of gluino production is the production and decay of squarks, particularly through the process $q\overline{q}\rightarrow \tilde{q}\overline{\tilde{q}}$ mediated by t-channel gluino exchange. The (anti-) squarks decay essentially instantly to a gluino and an (anti-) quark\footnote{We assume $\Delta m = 1$ GeV for definiteness.  There is no symmetry enforcing $m_{\tilde{g}} = m_{sq}$, so consideration of smaller mass splittings (which require careful treatment of non-perturbative QCD) is not motivated. }.  In this limit the (anti-) quark is soft and does not create a jet that could mask the R-hadron track and the gluino has the same kinematics as its parent squark.  In this regime, all coloured SUSY particles can simply be replaced by a gluino for the purpose of this analysis and we take the cross-section for gluino pair production to be the summed cross-section to any pair of SUSY coloured states.
\item {\it Squark mass agnostic}. When squarks are sufficiently heavier than the gluino, the quark emitted in the squark decay forms a jet which may obscure detection of the gluino track, thus reducing the sensitivity to gluinos produced via squark decay.  The most conservative limit for an arbitrary squark mass, which can be given without detailed consideration of event properties, is therefore obtained by considering only directly produced gluino pairs.   We simplify the parameter space by taking the light squark masses to be degenerate\footnote{To an excellent approximation, the gluino pair production cross-section depends on the mass of the up and down squarks only. The up and down squarks are typically degenerate at the electroweak scale, so the assumption of a common squark mass is well founded.}, then use the squark mass value that minimizes gluino pair production; this occurs when the squarks are approximately degenerate with the gluino.
\end{enumerate}

{\bf Gluino hadronization} \\
Monte Carlo samples of the final states were generated for a $\sqrt{s} = 1.96$ TeV
$p\bar{p}$ collider using Pythia \cite{Sjostrand:2006za}. The squarks
and gluinos are hadronized using the same hadronization scheme as in
previous studies \cite{Mackeprang:2009ad,Mackeprang:2006gx}.
Hadronization is restricted to $R$-mesons only, since $R$-baryon
production should be suppressed in a similar way that ordinary baryon
production is suppressed compared to meson production. As the
proportion of baryons produced compared to all particles is small in data
($<5$\%), and we expect a similar ratio for $R$-baryons, this will be counted
as a systematic error on the cross section.  The gluinoball formation
probability is set to 10\%{} compared to the $R$-mesons.

{\bf $R$-hadron propagation} \\
This sample of gluinoballs and $R$-mesons is then injected and propagated through the detector. For this work, a model of $R$-hadron scattering implemented in {\sc Geant-4}~\cite{Agostinelli:2002hh} is used~\cite{Mackeprang:2006gx}. 
In view of the inherent uncertainties associated with modeling $R$-hadron scattering, a pragmatic approach
based on analogy with observed low energy hadron scattering is adopted.  The scattering rate is estimated using a constant geometric cross-section of 12~mb per light ($u,d$) quark and 6~mb per strange quark, all 2-to-2 and 2-to-3 processes are allowed if they are kinematically feasible, and charge
conservation is respected. The proportion of 2-to-2 and 2-to-3 reactions is governed by phase space factors; no explicit constraints are applied to the probability of baryon number exchange. 

\begin{figure}[!ht]
  \epsfig{file=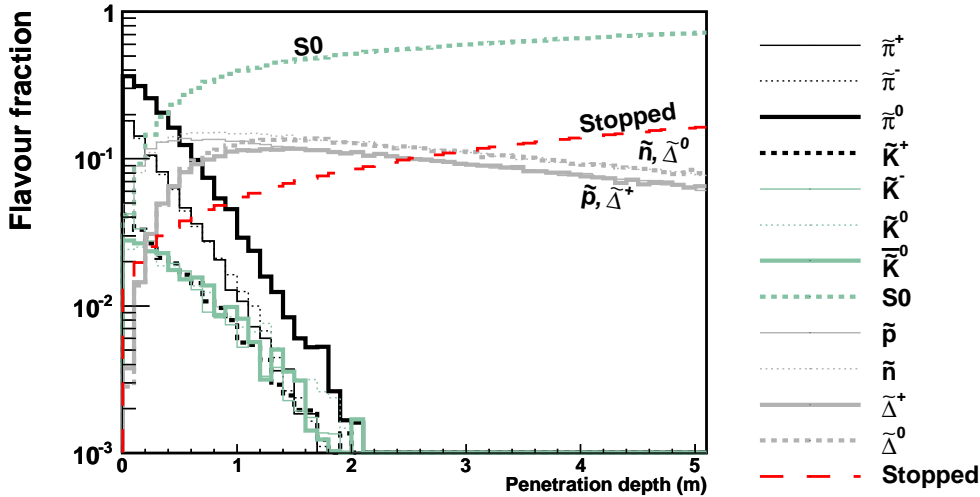,width=\textwidth}
  \caption{Flavour fractions for $R$-hadrons as a function of
    penetration depth in iron. The fractions are calculated from the
    percentage of gluino $R$-hadrons that have not stopped at the given
    depth. The stopping fraction is shown by the long dashed (red) line. We label the populations that are significant past 2m to aid in differentiating these lines.}\label{fig:flavfrac}
\end{figure}

Figure \ref{fig:flavfrac} shows the evolution of the flavour
composition of an $R$-hadron sample for a gluino mass of 200 GeV. A typical gluino hadronises to a meson immediately after production, converts to an R-baryon within the first meter or so of the detector, and then propagates through the calorimeter losing energy through electromagnetic interactions and changing charge through strong interactions with the valence quarks. Some fraction of the R-baryons stop in the detector, some enter the outer muon chamber charged, and the majority enter the muon chamber neutral. The
kinematic distribution of the $R$-hadrons is taken from a Pythia event
sample for this gluino mass.

\section{Re-evaluating the CDF CHAMP limit for the case of meta-stable gluinos}

\begin{figure}[!ht]
\centering
  \epsfig{file=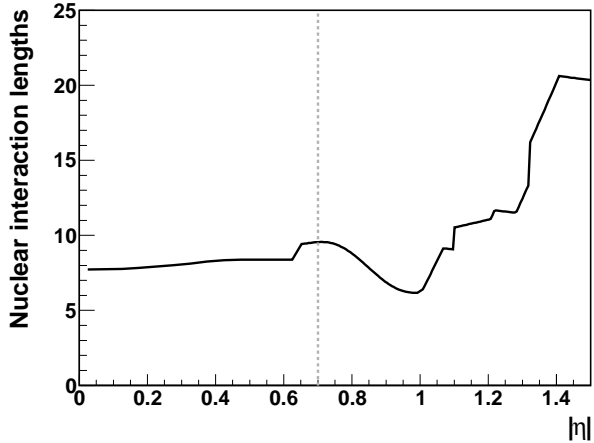,width=8cm}
  \caption{Matter distribution in the CDF detector as a function of pseudo-rapidity $| \eta |$.}\label{fig:cdf}
\end{figure}

In this section we apply the $R$-hadron production and propagation
models developed above to obtain a limit on the mass of a quasi-stable
gluino from the CDF Charge Massive Particle (CHAMP)
search~\cite{Aaltonen:2009kea}. The CDF detector is described in
detail in \cite{Blair:1996kx}. For the purpose of this paper, we will
be focusing on the number of nuclear interaction lengths between the
interaction point and the muon system. This is depicted in Figure
\ref{fig:cdf}. The pseudorapidity region considered in
\cite{Aaltonen:2009kea} is bounded by the grey dashed line.

\begin{figure}[!ht]
  \centering
  \epsfig{file=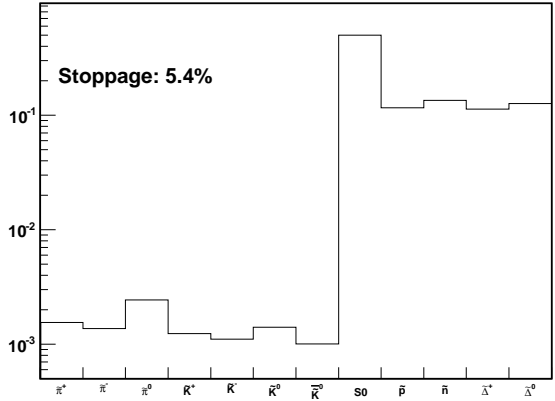,width=8cm}
  \caption{Flavour fractions for $R$-hadrons emerging from the simulated geometry. }\label{fig:flavcdf}
\end{figure}

To estimate the response of the CDF detector, the model described in
Sec. \ref{sec:sim} is supplemented with a sample of gluino $R$-hadrons
generated with Pythia \cite{Sjostrand:2006za}. Kinematic distributions
were derived from the sample, and high-statistics simulation samples
were generated obeying kinematic distributions extracted from
Pythia. The CDF detector was simulated using a depth of iron extracted
from Figure \ref{fig:cdf} at the generated pseudorapidity. The flavour
distribution of the emerging $R$-hadrons is shown in Figure
\ref{fig:flavcdf}.

To emulate the analysis performed in \cite{Aaltonen:2009kea}, only
$R$-hadrons
\begin{itemize}
\item in the pseudo-rapidity interval $|\eta|<0.7$ 
\item in the $\beta$ interval $0.4<\beta<0.9$
\item with a transverse momentum above $40$~GeV \emph{after traversal of the simulated iron}
\end{itemize}
were considered. Furthermore only $R$-hadrons that were positively
charged both immediately after hadronization and after traversing the
simulated geometry were considered. This satisfies the requirement in \cite{Aaltonen:2009kea} that the R-hadron leaves a charged track in the central tracker and the outer muon chambers.

The probability for an $R$-hadron to be accepted turns out to be of
order 10\%{}.  Denoting the single-object acceptance probability by
$x$, the resulting efficiency for pair-production events to be
accepted becomes
\begin{equation}
  \label{eq:eff}
  \epsilon_{\rm Vis} = 2x(1-x)+x^2
\end{equation}
yielding typical efficiencies of roughly 20\%{}.
\begin{table}[!ht]
  \centering
  \begin{tabular}{|lllll|}
    \hline
    Gluino mass: (GeV) & 200 & 300 & 400 & 500\\
    \hline
    \multicolumn{5}{|l|}{Visible Efficiency $\epsilon_{\rm Vis}$ (\%{})}\\
    \hline
    Base Model & 20.8 & 18.5 & 15.9 & 13.5\\
    \hline
    1. & 20.7 & 18.7 & 16.1 & 13.7 \\ % ZP
    2. & 23.6 & 21.3 & 18.5 & 15.7 \\ % Split
    3. & 26.4 & 22.6 & 19.0 & 15.8\\ %Split zero
    4. & 18.5 & 16.8 & 14.9 & 12.8 \\ %2 X xsec
    5. & 25.0 & 22.1 & 19.0 & 16.0 \\ %0.5 X xsec
    6. & 38.0 & 33.2 & 28.1 & 23.5 \\ % Regge
    \hline
  \end{tabular}
  \caption{$\epsilon_{\rm Vis}$ for gluino pair-production events calculated at four mass points. Rows 1-6 represent variations to the base model and are described in the text.}
  \label{tab:acc}
\end{table}
$\epsilon_{\rm Vis}$ is dependent upon details of the model for production, hadronisation and scattering. To check that our results are robust with respect to the theoretical uncertainty in these processes, we varied details of the model and studied the resulting change in $\epsilon_{\rm Vis}$. Table \ref{tab:acc} shows $\epsilon_{\rm Vis}$ calculated within the simulations for gluinos at four mass points. The model variations listed in the table are:
\begin{enumerate}
\item Increasing all  $R$-hadron masses by $1$~GeV. 
\item Increasing the $R$-meson/$R$-baryon mass splitting by $500$~MeV.
\item Setting the $R$-meson/$R$-baryon mass splitting to $0$~GeV.
\item Multiplying the nuclear scattering cross section by a factor of 2
\item Multiplying the nuclear scattering cross section by a factor of 0.5
\item Using the Regge model from ref. \cite{Mackeprang:2009ad} to
  model the nuclear scattering.
\end{enumerate}

In 1-3 we explore the sensitivity of the results to uncertainty in the spectrum of $R$-hadron masses, while 4-6 explore the sensitivity to uncertainty in the hadronic interactions of $R$-hadrons. The only variation that results in a lower visible fraction arises from doubling the hadronic scattering cross-section. All other model variations result in a larger visible fraction that would give a stronger limit on the mass of the gluino than the base model we take here. Thus the limits we obtain with this analysis are both conservative and robust against theoretical uncertainties\footnote{Our results are also robust with respect to the ordering of the R-baryon mass hierarchy. The R-baryons entering the muon chambers are 23\% charged to 77\% neutral with the default spectrum (see Fig. \ref{fig:flavcdf}), due to the large fraction of R-baryons that convert to the light stable $S^0$. If the $S^0$ were heavier, the R-baryons would be a democratic mixture of the lightest degenerate states  - giving approximately 50\% charged to 50\% neutral when they enter the muon chambers. A factor 2.2 increase in the charged fraction doubles the predicted signal improving the final limits to 423 GeV, 430 GeV, 345 GeV for the decoupled, degenerate and agnostic squark mass scenarios respectively. Therefore the light $S^0$ is both well motivated, and provides a conservative limit.}

\begin{figure}[!ht]
  \centering
  \epsfig{file=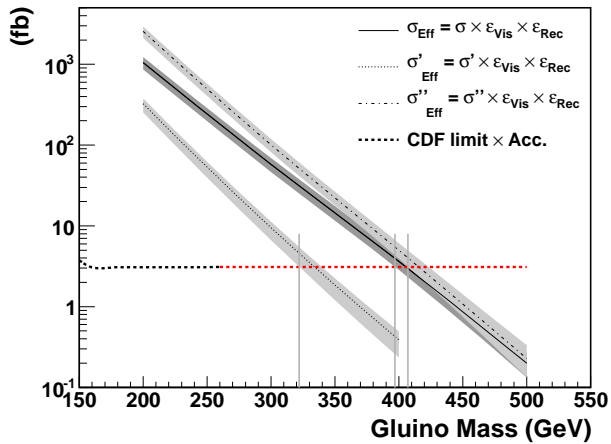,width=8cm}
  \caption{Calculated gluino pair production cross sections with efficiencies applied compared to the Tevatron pair production limit (horizontal dashed line). The solid line, $\sigma_{\rm Eff}$, corresponds to the split SUSY limit with decoupled squarks. The dotted line, $\sigma_{\rm Eff}'$, corresponds to the squark mass agnostic scenario. The dot-dashed line, $\sigma_{\rm Eff}''$, corresponds to the scenario with squarks degenerate with the gluino, in which the gluinos are primarily formed through the decays of (anti-)squarks.}\label{fig:limit}
\end{figure}

The result of the simulation is shown in Figure \ref{fig:limit}. The nearly horizontal line gives CDF's 95\%{}CL upper limit on stop squark pair production multiplied by the acceptance from \cite{Aaltonen:2009kea}.  The dependence of the acceptance on stop mass is virtually negligible, so can conservatively be extrapolated to the higher masses relevant for our limit as a constant.  The red line represents a constant extrapolation.

The three lines correspond to the three regimes outlined in section \ref{sec:sim}. In each case the cross-section is multiplied by the efficiency factors:
\begin{itemize}
\item \emph{Acceptance $\epsilon_{\rm Vis}$:}  The application of the nominal acceptance
  factors from Table \ref{tab:acc}.
\item \emph{Reconstruction efficiency $\epsilon_{\rm Rec}$:} A blanket reconstruction
  efficiency of 38 \%{} is applied in accordance with
  \cite{Aaltonen:2009kea}.
\end{itemize}
The solid line, $\sigma_{\rm Eff}$, corresponds to the split SUSY limit with decoupled squarks. The dotted line, $\sigma_{\rm Eff}'$, corresponds to the squark mass agnostic scenario. The dot-dashed line, $\sigma_{\rm Eff}''$, corresponds to the scenario with squarks degenerate with the gluino, in which the gluinos are primarily formed through the decays of (anti-)squarks.

The grey bands represent the systematic uncertainties. To calculate the error the renormalization and factorization scales were increased and decreased
by a factor of 2 and the standard CTEQ6 pdf~\cite{Nadolsky:2008zw}
was replaced by the MSTW pdf~\cite{Martin:2009iq}. The uncertainty is dominated largely by the scale
variation, and the error band represents the sum in quadrature of the
two sources of error. This pQCD systematic uncertainty far exceeds the
uncertainties quoted in Table \ref{tab:acc} and the latter are omitted in Figure \ref{fig:limit}.

We take the mass where the 95 \%{}CL Tevatron limit crosses the lower error band of the effective cross section $\sigma_{\rm Eff}$ to be a lower limit on the mass of a stable gluino. This gives lower limit on the gluino mass of $397$~GeV for decoupled squarks, $407$~GeV for squarks degenerate with the gluino, and $322$~GeV in the squark mass agnostic case.

\section{Conclusions}
\label{sec:concl}

We have used the results of the search for stop squark $R$-hadrons by CDF
\cite{Aaltonen:2009kea} to derive a limit on the production of a gluino-containing
$R$-hadron which traverses the detector without decaying, and hence a lower limit on the mass of a gluino with $\tau \gtrsim 30$ ns. Starting from QCD predictions for the
$R$-hadron mass spectra, we model the production, hadronisation and
scattering of $R$-hadrons within the CDF detector and calculate the
fraction of $R$-hadrons that mimic the signal investigated in
\cite{Aaltonen:2009kea}. We estimate the effects of the uncertainties
in the mass spectrum, gluino production cross section and details of the
scattering model. After including systematic uncertainties, we find a lower
limit of $407 \, {\rm GeV}$ on the mass of a quasi-stable gluino whose mass is approximately degenerate with squarks, a limit of $397 \, {\rm GeV}$ on the gluino mass if squarks are much more massive than gluinos, and a limit of $322$~GeV that is valid for any (degenerate) squark mass;  gluino mass limits can readily be derived for other squark mass scenarios. These are currently the strongest experimental limits on the mass of a stable or long-lived gluino and show the continued strength of
Tevatron data in constraining new physics.  This work also lays a foundation for interpreting future LHC searches.

\section*{Acknowledgments}

This research of GRF has been supported in part by NSF-PHY-0701451.  The work of JPR was supported by NSF Awards PHY-0758032, PHY-0449818 and NSF PHY-0900631, and DoE Award No. DE-FG02-06ER41417; he thanks the
Niels Bohr institute for their hospitality during the completion of this work. The work of RM was supported by the Danish National Research Foundation.

\appendix
\section{List of hadronic processes}\label{app:lists}
This appendix contains the full process lists used in this paper. No
distinction is used between $R_{8-0}$ and $R_{8-1}$ $R$-baryon states in
this list, excluding processes that specifically mix the states. It is
understood that any process written down for $R_{8-0}$ particles is
also valid for $R_{8-1}$.

\begin{table}[!htbp]
  \centering
  \begin{tabular}[t]{|>{$}l<{$}|}
    \hline
    R_{uds} + N \rightarrow R_{uds} + N\\
    \hline
    \slam+n\rightarrow\slam+n \\
    \slam+p\rightarrow\slam+p \\
    \hline
    R_{uds} + N \rightarrow R_{uds} + N + \pi:\\
    \hline
    \slam+n\rightarrow\slam+n + \pin \\
    \slam+n\rightarrow\slam+p + \pim \\
    \slam+p\rightarrow\slam+p + \pin \\
    \slam+p\rightarrow\slam+n + \pim \\
    \hline
    R_{uds}+N\rightarrow R_{8-0} +K +N\\
    \hline
    \slam + n \rightarrow \sneu + \knb + n\\
    \slam + n \rightarrow \sneu + \km + p\\
    \slam + n \rightarrow \spro + \km + n\\
    \slam + p \rightarrow \spro + \knb + n\\
    \slam + p \rightarrow \spro + \km + p\\
    \slam + p \rightarrow \sneu + \knb + p\\
    \hline
%  \end{tabular}
%  \begin{tabular}[t]{|>{$}l<{$}|}
%    \hline
    R_{8-0} + N \rightarrow R_{8-0} + N\\
    \hline
    \sneu + n \rightarrow \sneu + n\\
    \sneu + p \rightarrow \sneu + p\\
    \sneu + p \rightarrow \spro + n\\
    \spro + n \rightarrow \spro + n\\
    \spro + n \rightarrow \sneu + p\\
    \spro + p \rightarrow \spro + p\\
    \hline
    R_{8-0} + N \rightarrow R_{8-0} + N + \pi\\
    \hline
    \sneu + n \rightarrow \sneu + n + \pin\\
    \sneu + n \rightarrow \sneu + p + \pim\\
    \sneu + n \rightarrow \spro + n + \pim\\
    \sneu + p \rightarrow \sneu + p + \pin\\
    \sneu + p \rightarrow \sneu + n + \pip\\
    \sneu + p \rightarrow \spro + n + \pin\\
    \sneu + p \rightarrow \spro + p + \pim\\
    \spro + n \rightarrow \spro + n + \pin\\
    \spro + n \rightarrow \sneu + p + \pin\\
    \spro + n \rightarrow \sneu + n + \pip\\
    \spro + n \rightarrow \spro + p + \pim\\
    \spro + p \rightarrow \spro + p + \pin\\
    \spro + p \rightarrow \spro + n + \pip\\
    \spro + p \rightarrow \sneu + p + \pip\\
    \hline
  \end{tabular}
  \begin{tabular}[t]{|>{$}l<{$}|}
    \hline
    R_{8-0} + N \rightarrow R_{8-1} + N\\
    \hline
    \sneu + n \rightarrow \deln + n\\
    \sneu + p \rightarrow \deln + p\\
    \sneu + p \rightarrow \delp + n\\
    \spro + n \rightarrow \delp + n\\
    \spro + n \rightarrow \deln + p\\
    \spro + p \rightarrow \delp + p\\
    \hline
    R_{8-0} + N \rightarrow R_{8-1} + N + \pi\\
    \hline
    \sneu + n \rightarrow \deln + n + \pin\\
    \sneu + n \rightarrow \deln + p + \pim\\
    \sneu + n \rightarrow \delp + n + \pim\\
    \sneu + p \rightarrow \deln + p + \pin\\
    \sneu + p \rightarrow \deln + n + \pip\\
    \sneu + p \rightarrow \delp + n + \pin\\
    \sneu + p \rightarrow \delp + p + \pim\\
    \spro + n \rightarrow \delp + n + \pin\\
    \spro + n \rightarrow \deln + p + \pin\\
    \spro + n \rightarrow \deln + n + \pip\\
    \spro + n \rightarrow \delp + p + \pim\\
    \spro + p \rightarrow \delp + p + \pin\\
    \spro + p \rightarrow \delp + n + \pip\\
    \spro + p \rightarrow \deln + p + \pip\\
    \hline
    R_{8-0} + N \rightarrow R_{uds} + K + N\\
    \hline
    \sneu + n \rightarrow \slam + \kn + n\\
    \sneu + p \rightarrow \slam + \kn + p\\
    \sneu + p \rightarrow \slam + \kp + n\\
    \spro + n \rightarrow \slam + \kn + p\\
    \spro + n \rightarrow \slam + \kp + n\\
    \spro + p \rightarrow \slam + \kp + p\\
    \hline
  \end{tabular}
  \caption{Process lists galore for baryons}
  \label{tab:plists}
\end{table}

\begin{table}[!htbp]
  \centering
  \begin{tabular}[t]{|>{$}l<{$}|}
    \hline
    \spi + N \rightarrow \spi + N\\
    \hline
    \spin + n \rightarrow \spin + n\\
    \spin + n \rightarrow \spim + p\\
    \spin + p \rightarrow \spin + p\\
    \spin + p \rightarrow \spip + n\\
    \spim + n \rightarrow \spim + n\\
    \spim + p \rightarrow \spim + p\\
    \spim + p \rightarrow \spin + n\\
    \spip + n \rightarrow \spip + n\\
    \spip + n \rightarrow \spin + p\\
    \spip + p \rightarrow \spip + p\\
    \hline
    \spi + N \rightarrow \spi + N + \pi\\
    \hline
    \spin + n \rightarrow \spin + n + \pin\\
    \spin + n \rightarrow \spin + p + \pim\\
    \spin + n \rightarrow \spip + n + \pim\\
    \spin + n \rightarrow \spim + p + \pin\\
    \spin + n \rightarrow \spim + n + \pip\\
    \spin + p \rightarrow \spin + p + \pin\\
    \spin + p \rightarrow \spin + n + \pip\\
    \spin + p \rightarrow \spip + n + \pin\\
    \spin + p \rightarrow \spip + p + \pim\\
    \spin + p \rightarrow \spim + p + \pip\\
    \hline
    \spi + N \rightarrow R_{8-0} + \pi\\
    \hline
    \spin + n \rightarrow \sneu + \pin\\
    \spin + n \rightarrow \spro + \pim\\
    \spin + p \rightarrow \sneu + \pip\\
    \spin + p \rightarrow \spro + \pin\\
    \spip + n \rightarrow \spro + \pin\\
    \spip + n \rightarrow \sneu + \pip\\
    \spip + p \rightarrow \spro + \pip\\
    \spim + n \rightarrow \sneu + \pim\\
    \spim + p \rightarrow \sneu + \pin\\
    \spim + p \rightarrow \spro + \pim\\
    \hline
  \end{tabular}
  \begin{tabular}[t]{|>{$}l<{$}|}
    \hline
    \spi + N \rightarrow R_{8-0} + \pi + \pi\\
    \hline
    \spin + n \rightarrow \sneu + \pin +\pin\\
    \spin + n \rightarrow \sneu + \pip +\pim\\
    \spin + n \rightarrow \spro + \pim +\pin\\
    \spin + p \rightarrow \sneu + \pip +\pin\\
    \spin + p \rightarrow \spro + \pin +\pin\\
    \spin + p \rightarrow \spro + \pip +\pim\\
    \spip + n \rightarrow \spro + \pin +\pin\\
    \spip + n \rightarrow \spro + \pip +\pim\\
    \spip + n \rightarrow \sneu + \pip +\pin\\
    \spip + p \rightarrow \spro + \pip +\pim\\
    \spip + p \rightarrow \sneu + \pip +\pip\\
    \spim + n \rightarrow \sneu + \pim +\pin\\
    \spim + n \rightarrow \spro + \pim +\pim\\
    \spim + p \rightarrow \sneu + \pin +\pin\\
    \spim + p \rightarrow \sneu + \pip +\pim\\
    \spim + p \rightarrow \spro + \pim +\pin\\
    \hline
    \spi + N \rightarrow R_{uds} + K\\
    \hline
    \spin + n \rightarrow \slam + \kn \\
    \spin + p \rightarrow \slam + \kp \\
    \spip + n \rightarrow \slam + \kp \\
    \spim + p \rightarrow \slam + \kn \\
    \hline
    \spi + N \rightarrow R_{uds} + K + \pi\\
    \hline
    \spin + n \rightarrow \slam + \kn + \pin \\
    \spin + n \rightarrow \slam + \kp + \pim \\
    \spin + p \rightarrow \slam + \kp + \pin \\
    \spin + p \rightarrow \slam + \kn + \pip \\
    \spip + n \rightarrow \slam + \kp + \pin \\
    \spip + n \rightarrow \slam + \kn + \pip \\
    \spip + p \rightarrow \slam + \kp + \pip \\
    \spim + n \rightarrow \slam + \kn + \pim \\
    \spim + p \rightarrow \slam + \kn + \pin \\
    \spim + p \rightarrow \slam + \kp + \pim \\
    \hline
  \end{tabular}
  \begin{tabular}[t]{|>{$}l<{$}|}
    \hline
    \ska + N \rightarrow \ska + N\\
    \hline
    \skap + n \rightarrow \skap + n\\
    \skap + n \rightarrow \skan + p\\
    \skap + p \rightarrow \skap + p\\
    \skam + n \rightarrow \skam + n\\
    \skam + p \rightarrow \skam + p\\
    \skam + p \rightarrow \skanb + n\\
    \skan + n \rightarrow \skan + n\\
    \skan + p \rightarrow \skan + p\\
    \skan + p \rightarrow \skap + n\\
    \skanb + n \rightarrow \skanb + n\\
    \skanb + n \rightarrow \skam + p\\
    \skanb + p \rightarrow \skanb + p\\
    \hline
    \ska + N \rightarrow \ska + N + \pi\\
    \hline
    \skap + n \rightarrow \skap + n + \pin\\
    \skap + n \rightarrow \skan + p + \pin\\
    \skap + n \rightarrow \skan + n + \pip\\
    \skap + n \rightarrow \skap + p + \pim\\
    \skap + p \rightarrow \skap + p + \pin\\
    \skap + p \rightarrow \skap + n + \pip\\
    \skap + p \rightarrow \skan + p + \pip\\
    \skam + n \rightarrow \skam + n + \pin\\
    \skam + n \rightarrow \skam + p + \pim\\
    \skam + n \rightarrow \skanb + n + \pim\\
    \skam + p \rightarrow \skam + p + \pin\\
    \skam + p \rightarrow \skam + n + \pip\\
    \skam + p \rightarrow \skanb + n + \pin\\
    \skam + p \rightarrow \skanb + p + \pim\\
    \skan + n \rightarrow \skan + n + \pin\\
    \skan + n \rightarrow \skan + p + \pim\\
    \skan + n \rightarrow \skap + n + \pim\\
    \skan + p \rightarrow \skan + p + \pin\\
    \skan + p \rightarrow \skan + n + \pip\\
    \skan + p \rightarrow \skap + n + \pin\\
    \skan + p \rightarrow \skap + p + \pim\\
    \skanb + n \rightarrow \skanb + n + \pin\\
    \skanb + n \rightarrow \skanb + p + \pim\\
    \skanb + n \rightarrow \skam + p + \pin\\
    \skanb + n \rightarrow \skam + n + \pip\\
    \skanb + p \rightarrow \skanb + p + \pin\\
    \skanb + p \rightarrow \skanb + n + \pip\\
    \skanb + p \rightarrow \skam + p + \pip\\
    \hline
  \end{tabular}
  \caption{Process lists galore for mesons}
  \label{tab:procmes}
\end{table}

\begin{table}[!htbp]
  \centering
  \begin{tabular}[t]{|>{$}l<{$}|}
    \hline
    \ska + N \rightarrow R_{uds} + \pi\\
    \hline
    \skam + n  \rightarrow \slam + \pim\\
    \skam + p  \rightarrow \slam + \pin\\
    \skanb + n \rightarrow \slam + \pin\\
    \skanb + p \rightarrow \slam + \pip\\
    \hline
    \ska + N \rightarrow R_{uds} + \pi + \pi\\
    \hline
    \skam + n  \rightarrow \slam + \pim + \pin\\
    \skam + p  \rightarrow \slam + \pin + \pin\\
    \skam + p  \rightarrow \slam + \pip + \pim\\
    \skanb + n  \rightarrow \slam + \pin + \pin\\
    \skanb + n  \rightarrow \slam + \pip + \pim\\
    \skanb + p  \rightarrow \slam + \pip + \pin\\
    \hline
    \ska + N \rightarrow R_{8-0} + K\\
    \hline
    \skap + n \rightarrow \sneu + \kp\\
    \skap + n \rightarrow \spro + \kn\\
    \skap + p \rightarrow \spro + \kp\\
    \skap + p \rightarrow \sneu + \kn\\
    \skam + n \rightarrow \sneu + \km\\
    \skam + p \rightarrow \spro + \km\\
    \skam + p \rightarrow \sneu + \knb\\
    \skan + n \rightarrow \sneu + \kn\\
    \skan + p \rightarrow \spro + \kn\\
    \skan + p \rightarrow \sneu + \kp\\
    \skanb + n \rightarrow \sneu + \knb\\
    \skanb + n \rightarrow \spro + \km\\
    \skanb + p \rightarrow \spro + \knb\\
    \hline
  \end{tabular}
  \begin{tabular}[t]{|>{$}l<{$}|}
    \hline
    \ska + N \rightarrow R_{8-0} + K + \pi\\
    \hline
    \skap + n \rightarrow \sneu + \kp + \pin\\
    \skap + n \rightarrow \sneu + \kn + \pip\\
    \skap + n \rightarrow \spro + \kn + \pin\\
    \skap + n \rightarrow \spro + \kp + \pim\\
    \skap + p \rightarrow \spro + \kp + \pin\\
    \skap + p \rightarrow \spro + \kn + \pip\\
    \skap + p \rightarrow \sneu + \kp + \pip\\
    \skam + n \rightarrow \sneu + \km + \pin\\
    \skam + n \rightarrow \sneu + \knb + \pim\\
    \skam + n \rightarrow \spro + \km + \pim\\
    \skam + p \rightarrow \spro + \km + \pin\\
    \skam + p \rightarrow \spro + \knb + \pim\\
    \skam + p \rightarrow \sneu + \knb + \pin\\
    \skam + p \rightarrow \sneu + \km + \pip\\
    \skanb + n \rightarrow \sneu + \knb + \pin\\
    \skanb + n \rightarrow \sneu + \km  + \pip\\
    \skanb + n \rightarrow \spro + \km  + \pin\\
    \skanb + n \rightarrow \spro + \knb + \pim\\
    \skanb + p \rightarrow \spro + \knb + \pin\\
    \skanb + p \rightarrow \spro + \km  + \pip\\
    \skanb + p \rightarrow \sneu + \knb + \pip\\
    \skan + n \rightarrow \sneu + \km  + \pin\\
    \skan + n \rightarrow \sneu + \knb + \pim\\
    \skan + n \rightarrow \spro + \km  + \pim\\
    \skan + p \rightarrow \spro + \km  + \pin\\
    \skan + p \rightarrow \spro + \knb + \pim\\
    \skan + p \rightarrow \sneu + \knb + \pin\\
    \skan + p \rightarrow \sneu + \km  + \pip\\
    \hline    
  \end{tabular}
  \caption{Baryon number changing processes for strange mesons}
  \label{tab:procska}
\end{table}


\begin{thebibliography}{99}

\bibitem{Fairbairn:2006gg}
  M.~Fairbairn, A.~C.~Kraan, D.~A.~Milstead, T.~Sj\"ostrand, P.~Skands and T.~Sloan,
  %``Stable massive particles at colliders,''
  Phys.\ Rept.\  {\bf 438} (2007) 1
  [arXiv:hep-ph/0611040].
  %%CITATION = PRPLC,438,1;%%
%\cite{Amsler:2008zzb}
\bibitem{Amsler:2008zzb}
  C.~Amsler {\it et al.} [ Particle Data Group Collaboration ],
  %``Review of Particle Physics,''
  Phys.\ Lett.\  {\bf B667 } (2008)  1.

%\cite{Farrar:1984gk}
\bibitem{Farrar:1984gk}
  G.~R.~Farrar,
  %``Light Gluinos,''
  Phys.\ Rev.\ Lett.\  {\bf 53 } (1984)  1029.
  
  %\cite{Heister:2003hc}
\bibitem{Heister:2003hc}
  A.~Heister {\it et al.} [ ALEPH Collaboration ],
  %``Search for stable hadronizing squarks and gluinos in e+ e- collisions up to s**(1/2) = 209-GeV,''
  Eur.\ Phys.\ J.\  {\bf C31 } (2003)  327-342.
  [hep-ex/0305071].
  
%\cite{Abdallah:2002qi}
\bibitem{Abdallah:2002qi}
  J.~Abdallah {\it et al.} [ DELPHI Collaboration ],
  %``Search for an LSP gluino at LEP with the DELPHI detector,''
  Eur.\ Phys.\ J.\  {\bf C26 } (2003)  505-525.
  [hep-ex/0303024].  
  
  %\cite{Aaltonen:2009kea}
\bibitem{Aaltonen:2009kea}
  T.~Aaltonen {\it et al.} [ CDF Collaboration ],
  %``Search for Long-Lived Massive Charged Particles in 1.96-TeV p anti-p Collisions,''
  Phys.\ Rev.\ Lett.\  {\bf 103 } (2009)  021802.
  [arXiv:0902.1266 [hep-ex]].  
%\cite{Aktas:2004pq}
  
  %\cite{Abazov:2007ht}
\bibitem{Abazov:2007ht}
  V.~M.~Abazov {\it et al.}  [D0 Collaboration],
  %``Search for stopped gluinos from $p\bar{p}$ collisions at $\sqrt{s}$ =
  %1.96-TeV,''
  Phys.\ Rev.\ Lett.\  {\bf 99}, 131801 (2007)
  [arXiv:0705.0306 [hep-ex]].
  %%CITATION = PRLTA,99,131801;%%
  
%\cite{Collaboration:2010uf}
\bibitem{Collaboration:2010uf}
  CMS~Collaboration,
  %``Search for Stopped Gluinos in pp collisions at sqrt s = 7 TeV,''
  arXiv:1011.5861 [hep-ex].
  %%CITATION = ARXIV:1011.5861;%%
  
\bibitem{Aktas:2004pq}
  A.~Aktas {\it et al.} [ H1 Collaboration ],
  %``Measurement of anti-deuteron photoproduction and a search for heavy stable charged particles at HERA,''
  Eur.\ Phys.\ J.\  {\bf C36 } (2004)  413-423.
  [hep-ex/0403056].

%\cite{Abazov:2008qu}
\bibitem{Abazov:2008qu}
  V.~M.~Abazov {\it et al.} [ D0 Collaboration ],
  %``Search for Long-Lived Charged Massive Particles with the D0 Detector,''
  Phys.\ Rev.\ Lett.\  {\bf 102 } (2009)  161802.
  [arXiv:0809.4472 [hep-ex]].  

%\cite{Kraan:2004tz}
\bibitem{Kraan:2004tz}
  A.~C.~Kraan,
  %``Interactions of heavy stable hadronizing particles,''
  Eur.\ Phys.\ J.\  {\bf C37 } (2004)  91-104.
  [hep-ex/0404001].
  
  %\cite{deBoer:2007ii}
\bibitem{deBoer:2007ii}
  Y.~R.~de Boer, A.~B.~Kaidalov, D.~A.~Milstead {\it et al.},
  %``Interactions of Heavy Hadrons using Regge Phenomenology and the Quark Gluon String Model,''
  J.\ Phys.\ G {\bf G35 } (2008)  075009.
  [arXiv:0710.3930 [hep-ph]].
  
  %\cite{Mackeprang:2009ad}
\bibitem{Mackeprang:2009ad}
  R.~Mackeprang, D.~Milstead,
  %``An Updated Description of Heavy-Hadron Interactions in GEANT-4,''
  Eur.\ Phys.\ J.\  {\bf C66 } (2010)  493-501.
  [arXiv:0908.1868 [hep-ph]].  


%\cite{Blair:1996kx}
\bibitem{Blair:1996kx}
  R.~Blair {\it et al.}  [CDF-II Collaboration],
  ``The CDF-II detector: Technical design report''.
  %%CITATION = FERMILAB-PUB-96-390-E;%%

%\cite{Sjostrand:2006za}
\bibitem{Sjostrand:2006za}
  T.~Sjostrand, S.~Mrenna and P.~Z.~Skands,
  %``PYTHIA 6.4 Physics and Manual,''
  JHEP {\bf 0605} (2006) 026
  [arXiv:hep-ph/0603175].
  %%CITATION = JHEPA,0605,026;%%

%\cite{Kraan:2005ji}
\bibitem{Kraan:2005ji}
  A.~C.~Kraan, J.~B.~Hansen and P.~Nevski,
  %``Discovery potential of $R$-hadrons with the ATLAS detector,''
  Eur.\ Phys.\ J.\  C {\bf 49}, 623 (2007)
  [arXiv:hep-ex/0511014].
  %%CITATION = EPHJA,C49,623;%%

%\cite{Mackeprang:2006gx}
\bibitem{Mackeprang:2006gx}
  R.~Mackeprang and A.~Rizzi,
  %``Interactions of coloured heavy stable particles in matter,''
  Eur.\ Phys.\ J.\  C {\bf 50} (2007) 353
  [arXiv:hep-ph/0612161].
  %%CITATION = EPHJA,C50,353;%%

%\cite{DeGrand:1975cf}
\bibitem{DeGrand:1975cf}
  T.~A.~DeGrand, R.~L.~Jaffe, K.~Johnson and J.~E.~Kiskis,
  %``Masses And Other Parameters Of The Light Hadrons,''
  Phys.\ Rev.\  D {\bf 12} (1975) 2060.
  %%CITATION = PHRVA,D12,2060;%%

%\cite{Chanowitz:1983ci}
\bibitem{Chanowitz}
  M.~S.~Chanowitz and S.~R.~Sharpe,
  %``Spectrum Of Gluino Bound States,''
  Phys.\ Lett.\  B {\bf 126}, 225 (1983).
  %%CITATION = PHLTA,B126,225;%%

%\cite{Buccella:1985cs}
\bibitem{Buccella}
  F.~Buccella, G.~R.~Farrar and A.~Pugliese,
  %``R Baryon Masses,''
  Phys.\ Lett.\  B {\bf 153}, 311 (1985).
  %%CITATION = PHLTA,B153,311;%%
  
  \bibitem{UKQCD-Rmasses}
  M.~Foster and C.~Michael  [UKQCD Collaboration],
  %``Hadrons with a heavy colour-adjoint particle,''
  Phys.\ Rev.\  D {\bf 59} (1999) 094509
  [arXiv:hep-lat/9811010].

\bibitem{Agostinelli:2002hh}
  S.~Agostinelli {\it et al.}  [GEANT4 Collaboration],
  %``GEANT4: A simulation toolkit,''
  Nucl.\ Instrum.\ Meth.\  A {\bf 506} (2003) 250.
  %%CITATION = NUIMA,A506,250;%% 

%\cite{Arvanitaki:2005nq}
\bibitem{Arvanitaki}
  A.~Arvanitaki, S.~Dimopoulos, A.~Pierce, S.~Rajendran and J.~G.~Wacker,
  %``Stopping gluinos,''
  Phys.\ Rev.\  D {\bf 76}, 055007 (2007)
  [arXiv:hep-ph/0506242].
  %%CITATION = PHRVA,D76,055007;%%

%\cite{Fairbairn:2006gg}
\bibitem{Fairbairn}
  M.~Fairbairn, A.~C.~Kraan, D.~A.~Milstead, T.~Sjostrand, P.~Skands and T.~Sloan,
  %``Stable massive particles at colliders,''
  Phys.\ Rept.\  {\bf 438}, 1 (2007)
  [arXiv:hep-ph/0611040].
  %%CITATION = PRPLC,438,1;%%

%\cite{Beenakker:1996ch}
\bibitem{Beenakker:1996ch}
  W.~Beenakker, R.~Hopker, M.~Spira and P.~M.~Zerwas,
  %``Squark and gluino production at hadron colliders,''
  Nucl.\ Phys.\  B {\bf 492} (1997) 51
  [arXiv:hep-ph/9610490].
  %%CITATION = NUPHA,B492,51;%%
  
  %\cite{Nadolsky:2008zw}
\bibitem{Nadolsky:2008zw}
  P.~M.~Nadolsky {\it et al.},
  %``Implications of CTEQ global analysis for collider observables,''
  Phys.\ Rev.\  D {\bf 78}, 013004 (2008)
  [arXiv:0802.0007 [hep-ph]].
  %%CITATION = PHRVA,D78,013004;%%

%\cite{Martin:2009iq}
\bibitem{Martin:2009iq}
  A.~D.~Martin, W.~J.~Stirling, R.~S.~Thorne and G.~Watt,
  %``Parton distributions for the LHC,''
  Eur.\ Phys.\ J.\  C {\bf 63}, 189 (2009)
  [arXiv:0901.0002 [hep-ph]].
  %%CITATION = EPHJA,C63,189;%%


\end{thebibliography}
\end{document}